\documentclass[aps,prper,reprint,amsmath,amssymb,floatfix]{revtex4-2}

\usepackage{graphicx}
\usepackage{hyperref}
\usepackage[utf8]{inputenc}
\usepackage[T1]{fontenc}
\usepackage{csquotes}

\begin{document}

\title{From Floors to Electrons: Using a Building Analogy and Cartooning to Teach Quantum Numbers}

\author{James Day}
\email{day@phas.ubc.ca}
\affiliation{Department of Physics and Astronomy, The University of British Columbia, Vancouver, BC, Canada}

\author{Simone Williamson}
\affiliation{Independent Illustrator}

\date{\today}

\begin{abstract}
Quantum numbers are abstract, unfamiliar, and difficult to anchor in everyday experience. We describe a cartoon analogy, developed through a scientist--artist collaboration, in which an atom is rendered as a residential building, an electron as an occupant, and the four quantum numbers as features of that building: floor ($n$), suite layout ($\ell$), suite orientation ($m_\ell$), and occupant handedness ($m_s$). The analogy is designed to give novices an organizing framework for the four quantum numbers as a layered ``address'' before the formalism arrives. We discuss strengths, deliberate breakdowns, and suggested classroom use, including a metacognitive activity in which students identify where the analogy fails. The full cartoon is included as supplemental material.
\end{abstract}

\maketitle

\section{Introduction}

Aspects of quantum physics are no longer confined to the upper years of a physics degree. Concepts like superposition or entanglement that were once reserved for second- or third-year undergraduate courses now deserve attention earlier in a student's curriculum. Technology is changing at a pace that requires engaged citizens to understand some of the quantum basics if they are to make sense of the world. This paper offers a cartoon building analogy that teachers can use to introduce quantum numbers to their students.

The difficulty with understanding quantum phenomena is that none of the subject matter maps onto intuitive, everyday experience. These learning difficulties are well-documented: secondary and introductory students struggle with probabilistic models, superposition, and abstract state descriptions \cite{krijtenburg2017}. For quantum numbers specifically, the challenge is that abstract states and their numerical addresses are hard to visualize without slipping into classical, deterministic pictures. Part of the solution comes from developing an analogy and a cartoon through a scientist--artist collaboration. The analogy uses an atom as a building, an electron as an occupant, and quantum numbers as building features. The full cartoon is included as supplemental material. The deliberate choice to use a cartoon medium is supported by the literature, which indicates that animation and comics are highly effective for physics instruction \cite{rogers2007}. Because cartoons are inherently less constrained by reality, they provide an ideal, low-anxiety medium for students to explore non-intuitive physical laws and increase engagement across diverse learner populations.

A cartoon analogy does two things at once: it organizes complexity and it gives students' intuition something concrete to grab onto, closing some of the distance between formalism and sense-making. The efficacy of an analogy is tied to its representational format \cite{podolefsky2006}; the visual and structural choices shape how students map from the familiar onto the abstract. By using an architectural representation, the cartoon deliberately constrains what students might infer and nudges them toward a hierarchical reading of quantum addresses.

Quantum numbers are a set of four values that describe the unique state, or ``address,'' of an electron in an atom, specifying its energy level, orbital shape, spatial orientation, and spin direction. Each electron has a unique set of these numbers, which dictate its location and behaviour within an atom. The pedagogical problem is that quantum numbers are abstract, unfamiliar, and hard to anchor in everyday experience.

Common strategies for introducing quantum numbers to the uninitiated include orbitals, probability clouds, and hydrogen atom models. Each carries built-in pedagogical traps. Orbitals give students a categorical structure, helping organize quantum numbers into a manageable taxonomy, and their shapes are memorable and give students something to hold onto. But these shapes get reified as objects, obscuring that orbitals are mathematical constructs, not physical things. Probability clouds emphasize that quantum mechanics is about distributions. They communicate uncertainty and avoid classical orbits (a genuine conceptual win). But they are visually rich and conceptually under-structured: students see ``fuzzy clouds'' without understanding what distinguishes different states. Quantum numbers can feel like labels pasted onto the fuzzy clouds rather than organizing principles that generate them. The hydrogen atom is analytically solvable and historically central, so it provides a clean, canonical example where quantum numbers arise naturally from symmetry and boundary conditions. But students easily overgeneralize, assuming that all atoms behave like hydrogen. This can delay or confuse later understanding of many-electron atoms, electron--electron interactions, and the limited physicality of single-electron wavefunctions.

All three of these strategies are expert representations. They are optimized for correctness and compactness, but not for novice sense-making. Students are shown finished objects (shapes, clouds, models) rather than being given a way to organize meaning as the ideas are introduced.

Our goal with this project was to create a visual analogy that anchors the concepts in familiar imagery without distorting the underlying physics (too much). The approach draws on the idea of ``analogical scaffolding,'' in which abstract physics ideas are built up through a sequence of coherent analogies rather than one-off comparisons \cite{podolefsky2007}. Here, the cartoon introduces quantum numbers sequentially as parts of a single address within a shared structure, so they do not land as four isolated labels. By favouring conceptual placement over visual realism, the cartoon gives novices a scaffold they can work from before the formalism arrives. Traditional approaches show students what quantum states look like; the cartoon helps them understand how quantum states are organized.

\section{Building the Analogy}

Through dialogue with colleagues, we have shaped the analogy so that the notion of an address is kept tied to the idea of quantum numbers as the coordinates for locating the electron within the atom. As the statisticians put it, ``all models are approximations. Essentially, all models are wrong, but some are useful'' \cite{box1987}. That stance fits our analogy exactly. The apartment building is ``wrong'' about the physics of an atom, but it gives novices an organizing framework to hang the ideas on before the mathematics arrives.

The principal quantum number, $n$, describes an electron's main energy level and represents its average distance from the nucleus. For this concept, the floor levels of a residential building are used. Novice students do not naturally connect energy with discrete, quantized levels, and they often conflate continuous and discrete energy pictures---for instance, reading spectral lines as absolute energies rather than as transitions between them \cite{escalada2004}. Building floors push back against that drift. They are strictly ordered, discretely separated, and mutually recognizable; an occupant cannot live between floors, which reinforces that ``higher floor'' means a discrete jump to ``higher energy.'' The floors in the analogy risk suggesting that energy levels are evenly spaced, so it is important to frame them as ordered but not uniformly separated: they encode ranking, not energetic distance. This limitation can be used productively as a bridge to deeper physics, such as why hydrogen and many-electron atoms differ, or why spectroscopic lines are inherently nonuniform.

The azimuthal quantum number, $\ell$, describes the shape of an electron's orbital and its subshell (indicating its angular momentum). For this, we use suite layout/type, which addresses a student's understanding of ``subshell'' as a qualitative category with a given $n$ (as opposed to just another ``level''). It is a classification of state type. This helps students to understand that, on a given floor, there are different kinds of units. Furthermore, it helps prevent them from thinking that $n$ is the whole story. One limitation is that, for the quantum number $\ell$, it can be any integer from 0 to $n-1$. In the real world, however, one will never find an increasing number of suite layouts as we move up to higher floors. Students often reify orbital shapes as physical objects, a persistent and serious pitfall in quantum pedagogy. Learners struggling with probabilistic abstractions instinctively try to turn the mathematics into something tangible \cite{ravaioli2020}. To counteract this reflex, instructors should frame orbitals not as physical containers for electrons, but as categories of allowed quantum states. This is akin to how a suite layout describes the kind of living arrangement possible without acting as the physical occupant itself. Orbital shapes arise from mathematics and symmetry, not from little objects in 3D space.

The magnetic quantum number, $m_\ell$, specifies the orientation of an electron's orbital in space and thereby its magnetic state within a subshell. For this, we use suite orientation. This addresses the issue that, once students accept ``types,'' they still need to understand that each type comes in multiple distinct variants. A west-facing unit will get the evening sun, and a north-facing unit (in the northern hemisphere, at least) might not get sufficient light for indoor plants. Orientation works because it creates multiple, equally valid instances of the same layout that are still meaningfully distinct. As before, a limitation of this aspect of the analogy is that, for quantum numbers, $m_\ell$ takes integer values from $-\ell$ to $+\ell$, determining the number of available orbitals in a subshell.

Finally, the spin quantum number, $m_s$, describes an electron's intrinsic angular momentum, or ``spin.'' For this, we use the occupant's handedness within the suite, as students need to grasp that one quantum label is an intrinsic two-valued tag that does not change the address but does change the occupant.\footnote{This was the most difficult choice of all four quantum numbers, but $m_s$ is a special case. We needed a binary option, and this number is different from the others in the sense that the first three quantum numbers define a location and this one is directly related to the electron itself. But we could not conceive of an occupant property that would preserve the notion of the Pauli Exclusion Principle. We opted for handedness because it at least carries some of the concept of spin (we use right- and left-hand rules all the time when dealing with cross products in physics). While our choice here was a struggle, we believe it is evidence that spin is categorically different---and that the analogy is doing real conceptual work by exposing that difference.}

\section{Classroom Use}

What originated as a public outreach project evolved into a formal pedagogy, through deliberate thought of how it could be used in the classroom. For grades 11--12 or a first-year modern physics course, the cartoon will work best when used to organize thinking rather than to replace formal instruction.

The cartoon works well as a tool for conceptual consolidation. Narrative-driven comics in physics classrooms have been shown to increase student motivation and help retention by packaging abstract ideas as connected visual sequences \cite{pathoni2020}. After the formal mathematical definitions of the four quantum numbers have been introduced, the instructor can present the cartoon and walk through how each visual element of the building corresponds to a component of the electron's `address' in the atom. This helps novices recognize that quantum numbers are not arbitrary labels but rather coordinates within a structured system.

The cartoon may also be used as a discussion tool during a lecture and alongside formal explanations. If embedded directly into lecture slides, then, as each symbol is defined, the corresponding element of the building can be highlighted, reinforcing how the four numbers work together to specify an electron's address. This allows the analogy to serve as a running visual reference as the instructor transitions to more conventional representations, such as orbitals, subshells, or electron configurations. This works because the analogy becomes a persistent visual anchor that students can return to as the mathematical description becomes more abstract.

Perhaps most powerfully, the cartoon can be used as a short activity in which the students test and extend the analogy. (All models are wrong, but some are useful.) This could be a reflective activity after the concept has been introduced. Students can be asked questions such as: Where does the analogy help you understand the quantum numbers? Where does it break down? Can you extend the analogy to include concepts such as electron configurations and the Pauli exclusion principle? This kind of discussion turns the cartoon into a tool for metacognition, prompting students to think about how models and analogies work in physics. This works because the students learn not just the concept itself but also how physicists use imperfect models to reason about unfamiliar systems.

\section{Strengths and Limitations of the Analogy}

Like all analogies, this one has real strengths and real limits, and being upfront about both is part of what makes it work. The main strength is structural: the building gives students a familiar hierarchy to hang the four quantum numbers on before the formalism arrives. Rather than encountering them as four disconnected labels, students see them as a layered address system (which is exactly what they are). Visual analogies of this kind work best when they map organization rather than physical behaviour, and quantum numbers are fundamentally organizational concepts \cite{podolefsky2007}.

The limits are equally worth naming. Floors suggest evenly spaced energy levels, which they are not. The number of suite types and orientations is fixed in a real building; quantum numbers are not. The handedness of the occupant is the weakest mapping of all, since spin is an intrinsic quantum property with no real classical counterpart. The visual representation of hands writing may lead students to think that spin is a spatial behaviour. Instructors can leverage this weakness to discuss intrinsic versus spatial properties. This weakness reflects a broader concern: students taught with classical analogies can do worse on topics like wave--particle duality and tunnelling than students taught without them, because classical intuitions about fixed trajectories, positions, and hard boundaries get in the way \cite{rodriguez2025}. The pedagogical remedy is not to avoid the analogy entirely, but to explicitly constrain it and deliberately break it. Making the breakdown explicit turns a limitation into a metacognitive exercise. The suggested classroom activity in which students find where the analogy fails turns this limitation into a feature: students who can identify where a model fails are already thinking like physicists.

\section{Conclusion}

Quantum numbers are abstract, unfamiliar, and difficult to anchor in everyday experience; students need a foothold before the formalism can do its work. The apartment building analogy, and the cartoon that grew from it, is one attempt to provide that foothold: a familiar structure that organizes the four main quantum numbers into a coherent address system before the mathematics takes over. What sharpened this analogy, compared with the early drafts, was the cross-disciplinary collaboration itself. Scientist--artist partnerships have already helped communicate phenomena such as quantum scattering \cite{day2022}, but they remain underused as a systematic way to develop physics pedagogy. Pairing a physicist with an illustrator who treats the page as a teaching surface yields representations that are both scientifically defensible and genuinely accessible. The analogy presented here is not the last word (all models are wrong, remember). We encourage readers to find better mappings, to extend the analogy in new directions, or discover productive ways to break it.

\bibliographystyle{apsrev4-2}
\bibliography{quantum_numbers}

\end{document}